\documentclass{ws-procs975x65}
\usepackage{graphicx}
\usepackage{xcolor}
\usepackage[normalem]{ulem}  

\def\beq{\begin{equation}}
\def\eeq{\end{equation}}

\begin{document}

\title{Noncongruence of phase transitions in strongly interacting matter}
\author{Matthias Hempel$^*$}
\address{Department of Physics, University of Basel, Basel, Switzerland\\
$^*$E-mail: matthias.hempel@unibas.ch}
\author{Veronica Dexheimer} 
\address{Department of Physics, Kent State University,Kent OH, USA}
\author{Stefan Schramm}
\address{FIAS, Johann Wolfgang Goethe University, Frankfurt, Germany}
\author{Igor Iosilevskiy}
\address{Joint Institute for High Temperature of RAS, Moscow, Russia \\
Moscow Institute of Physics \& Technology (State University), Moscow, Russia}

\begin{abstract}
First-order phase transitions (PTs) with more than one globally conserved charge, so-called noncongruent PTs, have characteristic differences compared to congruent PTs (e.g., dimensionality of phase diagrams and location of critical points and endpoints). Here we discuss the noncongruent features of the QCD PT and compare it with the nuclear liquid-gas (LG) PT, for symmetric and asymmetric matter in heavy-ion collisions and neutron stars. In addition, we have identified a principle difference between the LG and the QCD PT: they have opposite slopes in the pressure-temperature plane. 
\end{abstract}


\bodymatter

\section{Introduction}
It is well known that at low densities and temperatures there is the nuclear liquid-gas (LG) phase transition (PT), which is of first-order if Coulomb and surface interactions are neglected. At high temperatures and densities, a variety of different PTs have been postulated in the literature, which relate to different phases of quark matter. These PTs are typically presented in form of phase diagrams with baryon chemical potential $\mu_B$ and temperature $T$, or baryon number density $\rho_B$ and $T$, as state variables. In the $\mu_B$-$T$-plane, first-order PTs are typically expected to appear as lines, on which one has phase coexistence. It is well known that these PT lines turn into PT areas if they are shown in the $\rho_B$-$T$-plane. This is simply due to the fact that the densities of the two phases which are in phase coexistence are generally not the same. Phase coexistence at each temperature, therefore, extends over a range of densities, spanned by the densities of the two phases. It is also well known that there could be additional axes of the phase diagrams corresponding to additionally conserved or externally fixed quantities, such as isospin, strangeness, lepton fraction, or magnetic fields. In the present proceeding, which summarizes the results of Ref.~\refcite{hempel13}, we discuss some general effects of such additional state parameters on  qualitative features of phase diagrams. 

\section{Noncongruence}
A first-order PT with more than one globally conserved charge, with the value of the conserved charges being fixed/specified externally, is called a \textit{noncongruent} PT. A PT with only a single conserved charge is called a \textit{congruent} PT.\footnote{This definition is only valid for phase coexistence of two (or more) macroscopic phases without Coulomb interactions, see Refs.~\refcite{iosilevskiy10,hempel13}.} This terminology was introduced to the astrophysical and nuclear physics community in Ref.~\refcite{iosilevskiy10}. Noncongruent PTs have been studied thoroughly for chemically reacting plasmas in the context of nuclear reactor safety problems.\cite{I1,I4} The universal nature of this type of PT and its applicability for most astrophysical objects, e.g., for ``plasma'' PTs in the interiors of Jupiter and Saturn, brown dwarfs, etc., and for the quark-hadron PT in hybrid stars, was demonstrated in Refs.~\refcite{iosilevskiy03,iosilevskiy10}. The qualitative differences of congruent and noncongruent PTs will be discussed in Sec.~\ref{sec_results}. We point out that the importance of the number of conserved charges is already well known in the fields of heavy-ion collisions (HICs), nuclear physics and astrophysics, see, e.g., Refs.~\refcite{barranco80,greiner87,glendenning92,muller95,muller97}. Typically, the terms ``Maxwell'' and ``Gibbs'' are used to distinguish the two types of first-order PTs, whereas these terms mostly express the respective PT constructions that are used.

\section{General setup}
For the LGPT of nuclear matter, we employ the FSUgold model (Ref.~\refcite{todd2005}) and for the QCD PT the Chiral SU(3) model (Ref.~\refcite{Dexheimer:2009hi}). In all calculations we neglect Coulomb interactions and work with the ``Coulomb-less'' approximation. Depending on the external constraints that are imposed on the system (specified below), we solve the corresponding conditions for thermal, mechanical and chemical equilibrium.\cite{hempel09} 

\subsection{The FSUgold model}
is a relativistic mean-field model with nucleons as particle degrees of freedom. In addition to self-couplings of the sigma- and omega-mesons, a coupling between the omega- and the rho-meson is included to obtain a more realistic behavior of the symmetry energy and the neutron matter equation of state.\cite{todd2005}

\subsection{The Chiral SU(3) model}
(Ref.~\refcite{Dexheimer:2009hi}) is an effective relativistic mean-field model with interactions based on meson-exchange. It is consistent with various constraints from both astrophysics and nuclear experiments.\cite{dexheimer15} It is important for our purposes that both quarks (up, down, and strange) and hadrons (baryon octet) are included as particle degrees of freedom in a chemical mixture where quarks and hadrons coexist. Such a behavior is required if one wants to obtain a critical point and deconfinement in a unified EOS model.\cite{hempel13} The actual population of the considered degrees of freedom is regulated by the interactions.
   
\subsection{Considered scenarios}
First, we consider the QCD PT in neutron stars (NSs). In cold NSs, there is no conservation of strangeness and isospin and weak reactions have reached full equilibrium. The requirement of charge neutrality means that leptons have to be included. Within the Coulomb-less approximation, this can be realized in two different ways: by local charge neutrality (LCN) with $Y_Q^{\rm I}=Y_Q^{\rm II}=0$ or by global charge neutrality (GCN) $Y_Q=0$, where $Y_Q$ denotes the total electric charge per baryon. For LCN we have baryon number as the only globally conserved charge, for GCN there is electric charge as a second conserved charge. Thus, we expect that LCN results in a congruent and GCN in a noncongruent PT. 

Second, we investigate and compare the LG with the QCD PT for conditions of HICs. In HICs, there are at least two conserved charges: the (net) baryon number and the (net) isospin. Therefore, one can expect that PTs in HICs are generally noncongruent. Instead of isospin, it is equivalent to use $Y_Q$, with its value fixed by the charge and mass number of the colliding nuclei, $Y_Q=Z/A$. We consider two representative values: $Y_Q=0.5$ (isospin symmetric matter) and $Y_Q=0.3$ (isospin asymmetric matter). For simplicity here we suppress the strangeness degree of freedom by assuming zero net strangeness in each of the phases.

\section{Results and discussion}
\label{sec_results}
\subsection{QCD PT in NSs}
\begin{figure}
\begin{center}
\includegraphics[width=2in]{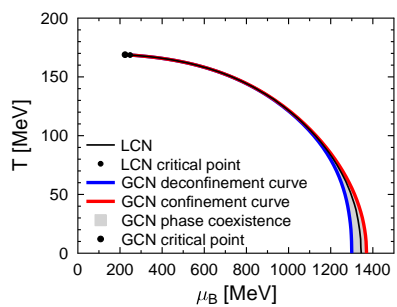}
\caption{Phase diagrams of the QCD PT for neutron star conditions for local (LCN) and global charge neutrality (GCN).}
\label{fig_pd_ns}
\end{center}
\end{figure}

The phase diagram for the QCD PT in NSs is shown in  Fig.~\ref{fig_pd_ns}. The assumption of GCN instead of LCN turns the (one-dimensional) PT line of a congruent PT into a (two-dimensional) PT region of a noncongruent PT. To distinguish the onset and end of the PT, we have introduced the deconfinement (DC) and confinement curves (CC). In the congruent case with LCN this distinction cannot be made in the $\mu_B$-$T$-plane, because the two lines coincide. For GCN, an isothermal PT occurs over a range of chemical potentials, corresponding also to a range of different pressures. In contrast, for LCN an isothermal PT occurs at a single chemical potential and pressure for each temperature. 

These results are common knowledge in the field of NSs, see, e.g., Ref.~\refcite{glendenning92}, but usually only $T=0$ is considered, corresponding to the case of cold NSs. In the following, we use this case as an example to explain the general features of noncongruent PTs: In a grand-canonical description, each phase corresponds to a hyper-surface in the intensive parameter space of $P$, $T$, $\mu_B$ and $\mu_Q$, where $\mu_Q$ is the charge chemical potential. Phase coexistence occurs, where the two hyper-surfaces intersect. For a nice graphical representation, see Ref.~\refcite{hanauske}. Consider a fixed constant temperature $T'$ and variable baryon chemical potential $\mu_B$. By imposing GCN, $Y_Q=0$, we have chosen a particular path through the parameter space $\mu_Q=\mu_Q(T',\mu_B,Y_Q=0)$ and $P=P(T',\mu_B,Y_Q=0)$. At the point where this path hits the intersection of the hyper-surfaces, corresponding to phase coexistence, generally the two phases have different baryon number densities \textit{and} different charge fractions. Because we impose a constant (global) charge fraction as an external constraint, there cannot be a direct transition into the other phase. By increasing $\mu_B$ further, the system remains instead on the intersection corresponding to phase coexistence, whereas $Y_Q^{\rm I}\neq0$ and $Y_Q^{\rm II} \neq 0$, but $Y_Q=0$. This implies that phase coexistence extends over a finite range of $\mu_B$, $P$, and $\mu_Q$.

The phase diagram shown in Fig.~\ref{fig_pd_ns} can thus be seen as a projection of the phase-coexistence region in the space of intensive variables $\{T,\mu_B,\mu_Q\}$ onto $\{T,\mu_B,Y_Q\}$ for the particular value of $Y_Q=0$. Note that there would be no qualitative effect of additional charges on the phase diagrams if, instead of their values, their corresponding chemical potentials were kept constant (e.g., by using $\mu_Q={\rm const.}$ instead of $Y_Q={\rm const.}$), see also Ref.~\refcite{ducoin06}. In this sense, the noncongruent features arise because we consider a ``mixed'' thermodynamic ensemble that is neither canonical nor grand-canonical.

We remark that there is another difference between congruent and noncongruent PTs, which is touched only briefly in the present proceeding: topological endpoints in the phase diagrams (e.g., the pressure and temperature endpoints) do not coincide with critical points for noncongruent PTs.\cite{iosilevskiy10,hempel13}

\subsection{LG and QCD PT in HICs}
\begin{figure}
\begin{center}
 \begin{tabular}{@{}cc@{}}
\includegraphics[width=2in]{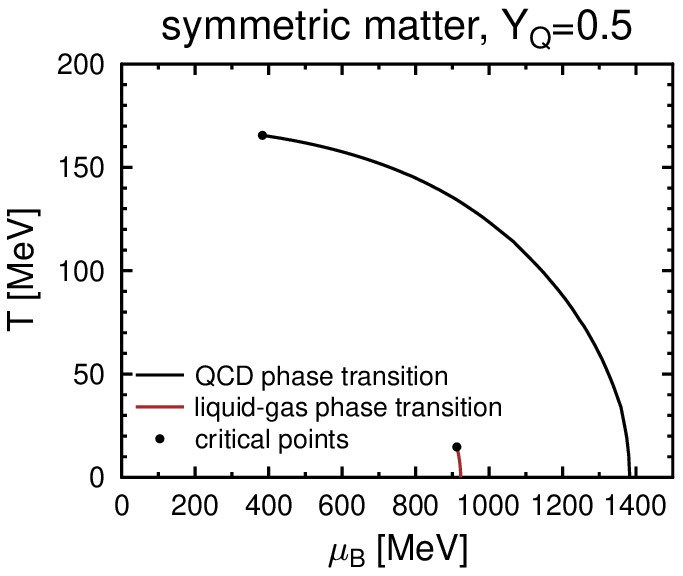} & 
\includegraphics[width=2in]{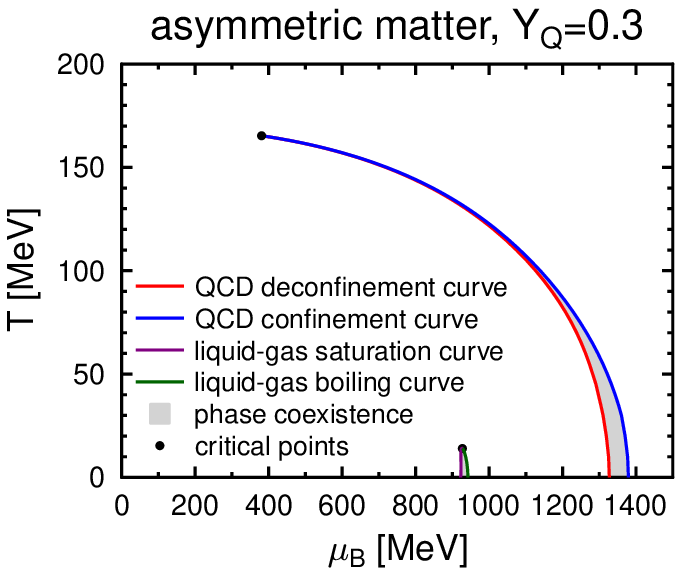} \\
\includegraphics[width=2in]{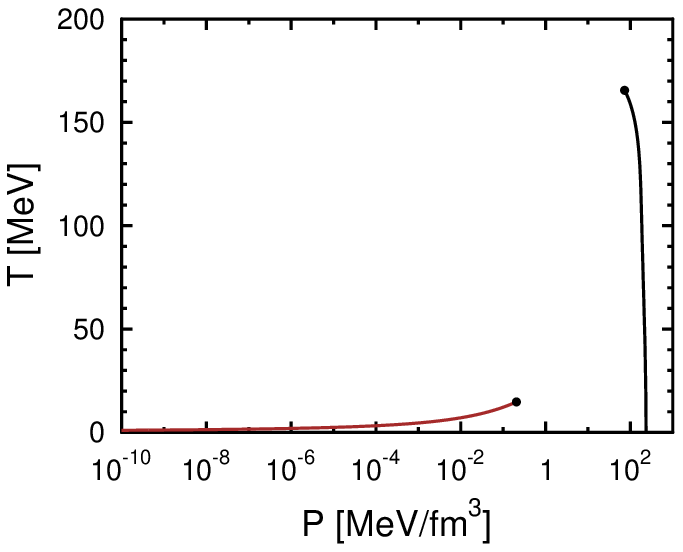} &
\includegraphics[width=2in]{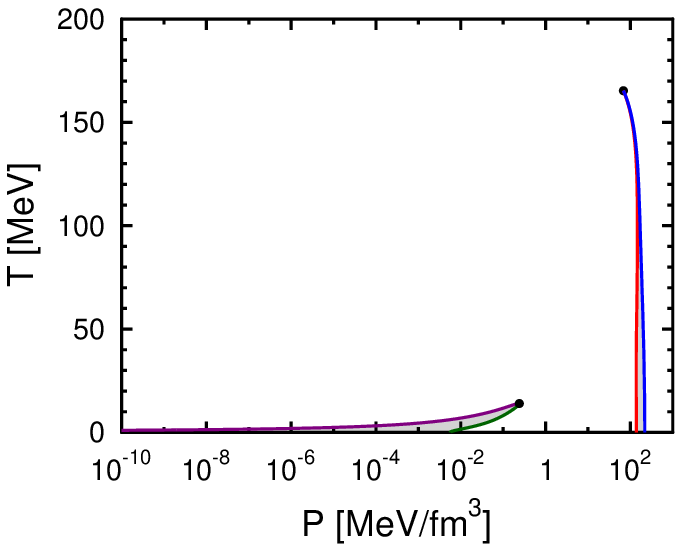} \\
\includegraphics[width=2in]{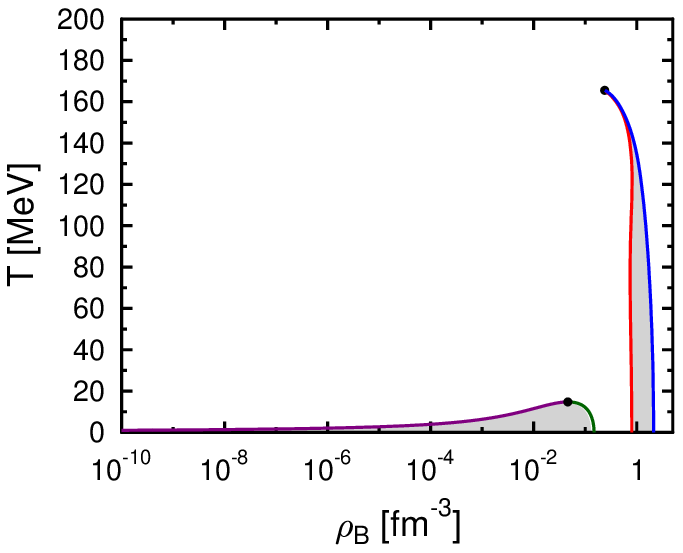} &
\includegraphics[width=2in]{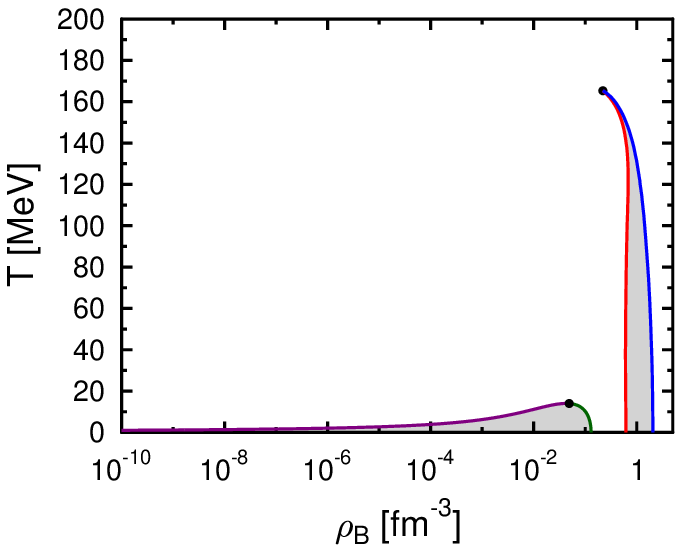}
\end{tabular}
\caption{Phase diagrams of the LG and QCD PTs. Left column: isospin symmetric matter, right column: isospin asymmetric matter.}
\label{fig_pd_general}
\end{center}
\end{figure}
In Fig.~\ref{fig_pd_general} we show the phase diagrams of isospin symmetric and isospin asymmetric matter, for both the LG and QCD PTs. The top left panel depicts the case of symmetric matter in the $\mu_B$-$T$-plane, where one has the well known behavior that the two PTs correspond to lines which terminate in a critical endpoint. In principle, also in this case we have isospin and baryon number as two independently conserved charges, but nevertheless the PTs are congruent. The reason lies in isospin-symmetry: independent of density and temperature, the isospin symmetric configuration has always the lowest free energy. Thus, also in phase coexistence both phases stay symmetric, which means that symmetric nuclear matter is an ``azeotrope''.\cite{muller95}

Conversely, in the isospin asymmetric case (top right panel) the system can gain energy by readjusting the isospin or charge fraction in the two phases. In consequence, the PT is noncongruent and visible as a two-dimensional phase coexistence region in the $\mu_B$-$T$-plane, equivalent to the case of GCN in NSs discussed above. For the QCD PT, one can identify an interesting property: around the critical point, the noncongruent features are vanishingly small, because the high temperatures reduce the importance of the values of the conserved charges.

In terms of density (two bottom panels of Fig.~\ref{fig_pd_general}), the phase diagrams of symmetric and asymmetric matter look qualitatively similar, in both cases one has two-dimensional phase coexistence regions. However, there is a difference that cannot be seen in the figure: the critical points of the noncongruent PTs in asymmetric matter do not coincide with the temperature endpoints as for the congruent PTs in symmetric matter, see, e.g., Refs.~\refcite{iosilevskiy10,hempel13}. The shapes of the phase diagrams in the $P$-$T$ plane (two middle panels) resemble the ones in the $\rho_B$-$T$ plane. However, because the pressure is an intensive quantity that has to be equal in coexisting phases, for the congruent PTs in symmetric matter one has PT lines. 

\subsection{Is the QCD PT of LG type?}
In the most common representation of the phase diagrams in high-energy physics, shown in the top panels of Fig.~\ref{fig_pd_general}, one gets the impression that the QCD and LGPTs are of the same kind, and that only the involved scales are different. However, it was demonstrated in Refs.~\refcite{iosilevskiy14,iosilevskiy15} that this impression is illusive. They belong to different subclasses of first-order PTs: the QCD PT is a typical \textit{entropic} PT, while the LGPT is a typical \textit{enthalpic} one. The main indicator of this discrimination is the opposite slope of the PT lines in the $P$-$T$ plane of the symmetric system, as can be seen in the middle left panel of Fig.~\ref{fig_pd_general}. This or other, related differences of the two PTs were noted also in Refs.~\refcite{satarov09,bombaci09,nakazato10,hempel13,steinheimer14}. According to the Clausius-Clapeyron equation, the sign of the slope $\left.\frac{dP}{dT}\right|_{\rm PT}$ is determined by the sign of the latent heat or, equivalently, by the difference of the specific entropy of the two phases. In the LGPT, one has the normal behavior that the denser phase (the liquid phase) has the lower specific entropy, leading to a positive latent heat. In the QCD PT, the opposite is the case, the latent heat is negative and the deconfined and chirally restored quark phase has the higher specific entropy due to the larger number of degrees of freedom (color and strangeness). 

In Refs.~\refcite{iosilevskiy14,iosilevskiy15} it was stressed and illustrated that the main consequence of entropic PTs are so-called ``abnormal'' thermodynamic properties in the phase coexistence region and potentially also in its close vicinity, where a great number of second cross derivatives of the thermodynamic potential have an unusual negative sign. An entropic QCD PT as found here can also lead to interesting consequences in astrophysics: Ref.~\refcite{yudin15} showed that it can induce a special form of ``inverted'' convection in proto-NSs, and Ref.~\refcite{hempel15} that it even can be related to the generation of core-collapse supernova explosions. 

\section*{Acknowledgments}
M.H.\ acknowledges support from the Swiss National Science Foundation (SNSF). Partial support comes from ``NewCompStar'', COST Action MP1304.
\bibliographystyle{ws-procs975x65}
\bibliography{references}

\end{document}